\documentclass[twocolumn, a4paper]{article}
\usepackage[T1]{fontenc}
\usepackage[utf8]{inputenc}
\usepackage{cite}
\usepackage{hyperref}
\usepackage{amsmath,amssymb,amsfonts}
\usepackage{algorithm}
\usepackage{lipsum}
\usepackage{authblk}
\usepackage[twocolumn,textwidth=17.31cm,columnsep=.81cm]{geometry}	
\newcommand\blfootnote[1]{%
  \begingroup
  \renewcommand\thefootnote{}\footnote{#1}%
  \addtocounter{footnote}{-1}%
  \endgroup
}
\usepackage[noend]{algpseudocode}
\makeatother
\usepackage{graphicx}
\usepackage{xcolor}

\title{Inter-frequency radio signal quality prediction for handover, evaluated in 3GPP LTE}

\author[$\dag$,$\ddag$]{Caroline Svahn}
\author[$\dag$]{Oleg Sysoev}
\author[$\ddag$]{Mirsad \u Cirki\'c}
\author[$\ddag$]{Fredrik Gunnarsson}
\author[$\ddag$]{Joel Berglund}

\affil[$\dag$]{\textit{Department of Computer and Information Science, Link\"oping University, Sweden.}}
\affil[$\ddag$]{\textit{Ericsson Research, Link\"oping, Sweden.}}

\date{}
\begin{document}

\maketitle
\blfootnote{This paper is a preprint. \textcopyright \,\ 2019 IEEE. Personal use of this material is permitted. Permission from IEEE must be obtained for all other uses, in any current or future media, including reprinting/republishing this material for advertising or promotional purposes,creating new collective works, for resale or redistribution to servers or lists, or reuse of any copyrighted component of this work in other works.}
\textbf{Radio resource management in cellular networks is typically based on device measurements reported to the serving base station. Frequent measuring of signal quality on available frequencies would allow for highly reliable networks and optimal connection at all times. However, these measurements are associated with costs, such as dedicated device time for performing measurements when the device will be unavailable for communication. To reduce the costs,  we consider predictions of inter-frequency radio quality measurements that are useful to assess potential inter-frequency handover decisions.  In this contribution, we have considered measurements from a live 3GPP LTE network. We demonstrate that straightforward applications of the most commonly used machine learning models are unable to provide high accuracy predictions. Instead, we propose a novel approach with a duo-threshold for high accuracy decision recommendations. Our approach leads to class specific prediction accuracies as high as 92\% and 95\%, still drastically reducing the need for inter-frequency measurements.}


\section{Introduction}
The booming interest in wireless services means that cellular networks are upgraded with service over multiple frequency carriers. This available resource is an asset, but also pose a relevant radio resource management problem of how to assign user connections to different frequency carriers, where users also are mobile. In order to support users in an adequate manner, the assignments need to be re-evaluated over time. If deemed necessary, user connection can be subject to handover \cite{handover} from one frequency carrier to another. To maintain relevant frequency assignments, frequent measurements would need to be performed, which would lead to unnecessary load of the network and extensive battery consumption. Signaling overhead can be  greatly reduced if an automatic prediction of the inter-frequency signal quality is provided. Predictions of the inter-frequency radio signal strength in terms of 3GPP LTE Reference Signal Receiver Power (RSRP) has been discussed in \cite{rsrp} by means of Random Forests. In this paper, we instead consider Reference Signal Receiver Quality (RSRQ) as this is a better measure of the actual network performance. Prediction of RSRQ is, however, more challenging than RSRP prediction as the RSRQ measure depends on the total received radio signal energy from all signals, Radio Signal Strength Indicator (RSSI), according to 

\begin{equation}
    RSRQ=\frac{R_b \cdot RSRP}{RSSI},
\end{equation}

\noindent where $R_b$ is the number of resource blocks \cite{rsrq}. RSSI depends on noise and interference, which makes RSRQ harder to predict, since the interference varies with the varying radio network load. 




In this paper, we predict inter-frequency RSRQ using the most commonly used machine learning methods: Random Forest, Neural Networks, Gaussian Processes and Logistic regression. We pose inter-frequency prediction as a classification problem. As our results demonstrate that a straightforward application of these methods leads to unsatisfactory prediction quality, we introduce a duo-threshold approach for decision recommendation. In this approach, we combine two phases: a phase in which we perform measurements of signal quality and a phase in which we deliver high-quality predictions by means of Random Forest models. \\
\indent The paper will be organized as follows: Section II will present Random Forest, Logistic regression, Gaussian Processes and Neural Networks as well as oversampling as a measure to counter unbalanced classes. In Section III we introduce our proposed duo-threshold approach. In Section IV we compare the algorithms in terms of performance and, finally, in Section V we summarize our findings.

\section{Models}
In this section, the methods used for RSRQ prediction are presented. Assume $m$ is the number of cells in the network. Let $P_i,\,\, i=1,\dots,m$ be an RSRP measurement from cell $i$ on the serving frequency and $Q_i,\,\, i=1,\dots,m$  be an RSRQ measurement from cell $i$ at the serving frequency. Let $j$ be the index of the serving cell, hence, $P_j$ and $Q_j$ are the corresponding measurements from the serving cell. Finally, denote the strongest RSRQ measurement on an alternative frequency as $y$. The feature vector in our models is then

\begin{equation}
    \mathbf{x}=(P_i, \dots, P_m,Q_i, \dots, Q_m).
\end{equation}

The aim of the classification is to model an ordering relationship of the type $y \geq Q_j$ as

\begin{equation}
    C= 
\begin{cases}
    1,& \text{if } y \leq Q_j\\
    2,& \text{if } Q_j < y.
\end{cases}
\label{eq:classes}
\end{equation}

\subsection{Random Forest}

The Random Forest is described in \cite{forest} as an ensemble of decision trees, where each tree is built by iterative splitting of the variable space into smaller sections, nodes, by simple conditions. A tree can be represented by a set of nodes, $\mathbf{M}=(M_j, \dots, M_J$). Each $M_j$ is assigned a value $V_j$ which is obtained by using a split variable $x_j$ and a split criterion $s_j$.  The model estimation is performed by applying Alg. \ref{rf}.

\begin{algorithm}[H]
\caption{Random Forest}\label{rf}
\begin{algorithmic}[1]
      \For{\texttt{$b=1$ to $B$}}
        \State Draw bootstrap sample \cite{boot}
        \State Grow random-forest tree $T_b$:
        \ForAll{nodes}
        \While{$m_{size}>m_{min}$}
         \State Randomly select $k$ variables
         \State Find best variable/split among them
         \State Split the node into two child nodes
         \EndWhile
         \EndFor
      \EndFor
\State Output the tree ensemble as $[T_b(\mathbf{x})]^B_1$,
\end{algorithmic}
\end{algorithm}

\noindent where $B$ is the number of bootstrap samples, $m_{size}$ is the node size and $m_{min}$ is the minimum node size. In our scenario, we use the default settings of $m_{min}=5$. The sample size of the bootstrap samples is $\lceil 0.632\cdot n_{obs}\rceil$, where $n_obs$ is the number of observations in the training set. $B=500$ trees are grown, and the best split is found by using the Gini impurity measure, see \cite{bishop} for details. We make predictions by locating a given $x$ in one of the terminal nodes.


Let $m_1$ be the number of observations of class 1 in a node in a single tree and $m_2$ the number of observations of class 2 in a node, then the output of a single classification tree is

\begin{equation}
    p(C=2|\mathbf{x})=\frac{m_2}{m_1+m_2},
\end{equation}

\noindent which is the probability of class $2$ being the majority class in the predicted node. The prediction of a Random Forest model is the average of predictions obtained from all trees in the ensemble.
\subsection{Logistic regression}



\indent Logistic regression is used for binary classification problems which requires transformation of the output to fit within the interval $[0,1]$ with the logistic sigmoid function. According to \cite{bishop}, for $M$ model parameters, we make new predictions with








\begin{equation}
    p(C=2|\mathbf{x})=\frac{1}{1+exp(-a)},
\end{equation}

\noindent where $a=\sum_{j=0}^{M-1}{\beta_j\mathbf{x}}$ and we find the weights, $\boldsymbol{\beta}$, with maximum likelihood using of the derivative of the logistic sigmoid function.

\subsection{Gaussian Processes}
A brief description of Gaussian Process classification is described in this section, however, details can be found in \cite{bishop}. We consider the model

\begin{equation}
    y(\mathbf{x})=\boldsymbol{\beta}^T\phi(\mathbf{x}),
\end{equation}

\noindent and search for the joint probability distribution of $y(x_1), \dots, y(x_N)$ by adding a prior distribution to the weight vector of the linear model with $\boldsymbol{\phi}=(\phi_0, \dots, \phi_{M-1})$ non-linear basis functions representing \textbf{x}:

\begin{equation}
    y(\mathbf{x},\beta)=\sum_{j=0}^{M-1}{\beta_j\phi_j(\mathbf{x})}.
\end{equation}

\noindent By using an isotropic Gaussian prior on $\boldsymbol{\beta}$ we get a posterior for $y$ from which we sample to obtain predictions. The posterior is Gaussian with zero-mean and covariance \textbf{K}, which is the so called Gram matrix from a kernel function. We run the Gaussian Processes using a vanilladot kernel.

\subsection{Neural Networks}
According to \cite{bishop}, when constructing a Neural Network, we compute linear combinations of $\mathbf{x}$ and arrange them to obtain $a_j$ \textit{activations}: 

\begin{equation}
    a_j=\sum^D_{i=1}\beta_{ji}x_{i}+\beta_{j0}.
\end{equation}

\noindent We perform transformations of the linear combinations, $z_j=h(a_j)$, and achieve \textit{hidden units}, where $h(\cdot)$ is, in our default settings, the logistic sigmoid function. To produce multiple hidden layers, we can again obtain linear combinations of the hidden units to form the next layer. To produce the output, the activations are transformed with the logistic sigmoid for our binary classification problem. The class probabilities are obtained with



\begin{equation}
     p(C=2|\mathbf{x})=\sigma\Big(\sum^M_{j=1}\beta_{kj}^{(2)}h\Big(\sum^D_{i=1}\beta_{ji}x_i^{(1)}+\beta_{j0}^{(1)}\Big)+\beta_{k0}^{(2)}\Big),
\end{equation}

\noindent where $\sigma$ is the logistic sigmoid function and the upper indices represent the layers.










\subsection{Oversampling}
As data can, in some cells, be rather imbalanced between the classes, oversampling of the minority class has been performed by randomly sampling from the minority class with replacement until reaching $50/50$ class proportions per cell/alternative frequency combination.
\section{Duo-threshold prediction}
 We propose a duo-threshold approach for which we either perform a measurement or deliver a high-quality prediction. True class values are computed by applying (\ref{eq:classes}). The overall accuracy is defined as $\frac{TP+TN}{TP+TN+FP+FN}$, where $TP$ is the number of positive observations predicted correctly, $FP$ the number of negative observations predicted as positive, $TN$ the number of negative observations predicted correctly and $FN$ the number of positive observations predicted as negative. Furthermore, we define observations for which the inter-frequency RSRQ is higher than the serving RSRQ as positive and observations for which the inter-frequency RSRQ is lower than the serving RSRQ as negative. In order to ensure high accuracy of the predictions, we introduce a three-class output with two thresholds, $\delta_1$ and $\delta_2$, as follows:

\begin{equation}
    C_d= 
\begin{cases}
    1,& \text{if } p(C=2|\mathbf{x})\leq \delta_1\\
    2,& \text{if } p(C=2|\mathbf{x})\geq \delta_2\\
    3,              & \text{otherwise},
\end{cases}
\label{eq:thclasses}
\end{equation}

\noindent where $p(C=2|\mathbf{x})$ is obtained from a machine learning model used for prediction. Equation (\ref{eq:thclasses}) demonstrates that observations not meeting any of the threshold conditions are classified into the third class as the predictions are too uncertain. With this approach, we can maintain a sufficient accuracy for the original classes as we only make decisions about the most certain observations. To compute the accuracies of the classes, we alter the traditional measurements True Positive Rate, $TPR=\frac{TP}{TP+FN}$, and True Negative Rate, $TNR= \frac{TN}{TN+FP}$. $TPR$ is the ratio of correctly classified true positives, therefore, $TPR$ corresponds to the accuracy of the observations where the alternative frequency has a higher RSRQ than the serving RSRQ and TNR to the accuracy of observations with a higher serving RSRQ. For our duo-threshold computations, we define $TNR_d$ and $TPR_d$ as

\begin{equation}
    TNR_d= \frac{TN+N_{mn}}{TN+FP+N_{mn}}
\end{equation}

\noindent and

\begin{equation}
    TPR_d= \frac{TP+N_{mp}}{TP+FN+N_{mp}}.
\end{equation}

\begin{figure}[htbp]
\centerline{\includegraphics[width=90mm]{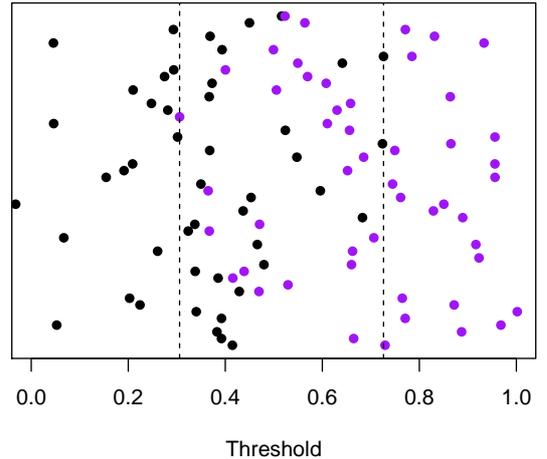}}
\caption{Example of the proposed threshold bounds.}
\label{fig}
\end{figure}

\noindent $N_{mn}$ and $N_{mp}$ are the number of observations from class 1 and class 2 which were predicted as class 3, respectively. Since we measure $N_{mn}$ and $N_{mp}$, we consider them correctly handled and therefore we add them to the nominators. This enables a natural comparison to the $100\%$ accuracy we would achieve by frequent measurements: when $\delta_1\rightarrow 0$ and $\delta_2\rightarrow 1$,  $TN+FP \rightarrow 0$ and $TP+FN \rightarrow 0$ and we measure all observations, which gives $TNR_d \rightarrow 1$ and $TPR_d \rightarrow 1$. Note that the $TNR_d$ is affected by $\delta_2$ and $TPR_d$ by $\delta_1$. A demonstration of the threshold impact can be seen in Fig. \ref{fig}, where the colors represent the two different classes and the dashed lines the threshold separating the more certain predictions from the less certain.

\noindent We will also use \textit{measure share} to evaluate our models, which we define as $\frac{N_{mn}+N_{mp}}{N_{tot}}$, where $N_{tot}$ is the total number of predicted observations.

\section{Results}
\subsection{Setup}\label{AA}
The data are collected from an urban scenario and consist of RSRP and RSRQ from the serving cell and close-by cells as well as inter-frequency RSRP and RSRQ measurements. All covariates have been standardised, and the models have been fitted independently by cell/inter-frequency combination since the distributions among cells and inter-frequencies vary and the enumeration of the nearby cells vary between serving cells. The data are sparse since few cells provide a signal for each observation. For each observation, there is a pair $(RSRP, RSRQ)$ of inter-frequency measurements which is believed to represent the strongest available signal on the alternative frequency. A small portion of the datasets, consisting of each cell/inter-frequency combination, are too small (typically less observations than covariates) and have been removed for more fair comparison of the models. The models have been fitted and evaluated with the training and test partition 75/25. We allow some prediction error and therefore we set an acceptance limit for the test $TNR$ to $0.95$ and $TPR$ to $0.90$. In the handover scenario, moving a user from a serving frequency with known signal quality to an alternative, possibly inferior frequency without signal confirmation is combined with higher risk than letting the user stay on the serving frequency. Therefore, we set a stricter limit for $TNR$. We have used R \cite{R} and packages \texttt{randomForest} \cite{rForest}, \texttt{kernlab} \cite{rGP} and \texttt{neuralnet} \cite{nn}. \\
\indent All models were run with default settings as stated in Section II. 




\subsection{Model evaluation}
In Tab. \ref{tab:tab1}, the overall test accuracies as well as the $TPR$ and the $TNR$ for Random Forests (RF), Logistic regression (LM), Gaussian Processes (GP) and Neural Networks (NN) are shown for a random $10\%$ uniform sample of the data. We show results for the NN with $(50, 30, 10)$ neurons as NN did not perform better with more neurons or layers. Furthermore, we have used early stopping to avoid overfitting. The table represents means of the results of the sampled data sets.

\begin{table}[htbp]
\caption{Accuracies}
\begin{center}
\begin{tabular}{lccc} \hline
\textbf{Model} & \textbf{Acc. class}  & \textbf{TPR} & \textbf{TNR} \\\hline
\textbf{RF} & 0.801  & 0.324 & 0.896 \\
\textbf{LM} & 0.792  & 0.415 & 0.872 \\
\textbf{GP} & 0.791  & 0.313 & 0.890 \\
\textbf{NN} & 0.780 & 0.415 & 0.845\\\hline
\end{tabular}
\label{tab:tab1}
\end{center}
\end{table}


\noindent From the results, we conclude that NN have the lowest $TNR$ and the lowest general accuracy. LM has the best $TPR$ and the RF performs evenly, and has the best $TNR$, which is the most crucial measure in our scenario since a model with low $TNR$ may lead to unnecessary handovers. RF and LM appear to be the most promising models.\\
\indent The results when using oversampling can be found in Tab. \ref{tab:over}.


\begin{table}[htbp]
\caption{Accuracies, oversampling}
\begin{center}
\begin{tabular}{lcccc} \hline
\textbf{Model} & \textbf{Acc. class}  & \textbf{TPR} & \textbf{TNR} \\\hline
\textbf{RF} & 0.798  & 0.880 & 0.779 \\
\textbf{LM} & 0.626  & 0.874 & 0.571 \\
\textbf{GP} & 0.743  & 0.856 & 0.721 \\
\textbf{NN} & 0.802  & 0.894 & 0.784 \\\hline
\end{tabular}
\label{tab:over}
\end{center}
\end{table}

\noindent The $TPR$ is higher for all models, however, the $TNR$ and overall accuracy are lower. With oversampling, the NN models achieve better results. The NN appear to outperform the other models in terms of accuracy, $TPR$ and $TNR$. RF achieve the second highest values for all measures.
\indent We conclude that oversampling improves our predictions, however, does not reach $TNR=0.95$ and $TPR=0.90$. Therefore, we continue to examine RF and NN with oversampling when introducing our custom duo-threshold method. \\

\subsection{Threshold evaluation}

\begin{figure}[htbp]
\centerline{\includegraphics[width=90mm]{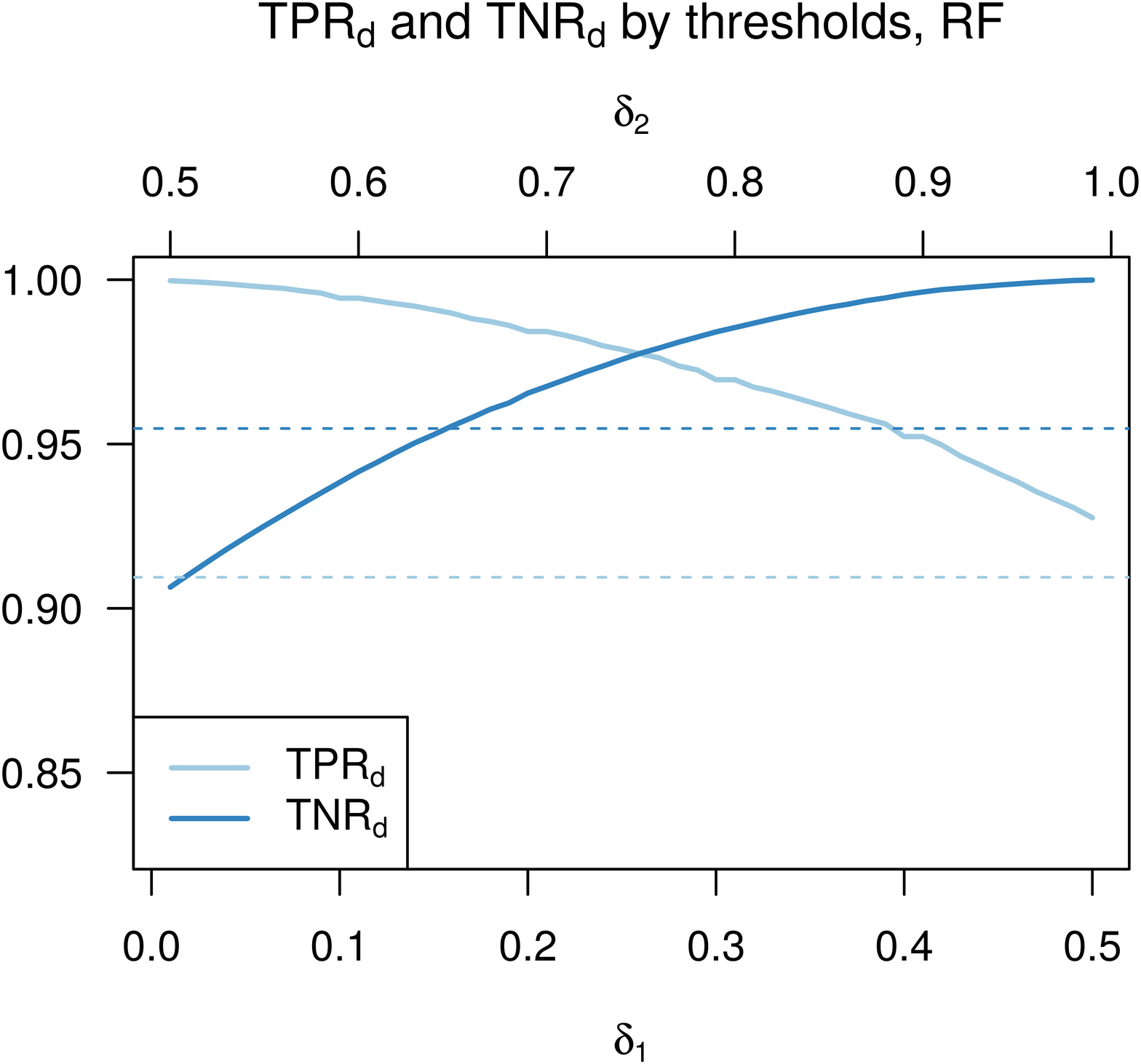}}
\caption{$TPR_d$ and $TNR_d$ for Random Forest by $\delta_1$ and $\delta_2$.}
\label{fignpvppvRF}
\end{figure}

\begin{figure}[htbp]
\centerline{\includegraphics[width=90mm]{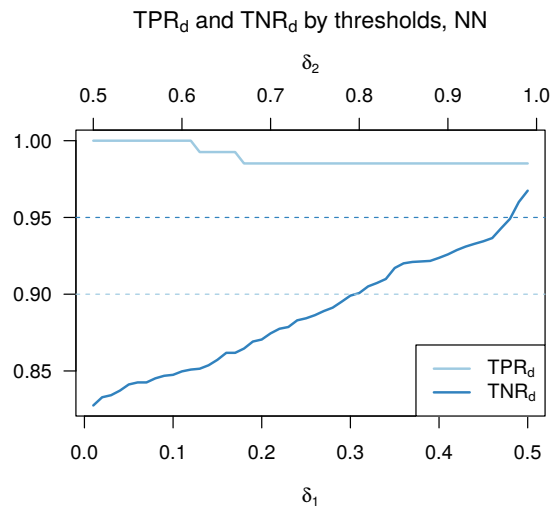}}
\caption{$TPR_d$ and $TNR_d$ for Neural Networks by $\delta_1$ and $\delta_2$.}
\label{fignpvppvLM}
\end{figure}

As the standard approaches do not provide satisfying results, a different strategy to deliver high-accuracy decisions is needed. We again set lower limits $TNR_d=0.95$ and $TPR_d=0.90$. The $TNR_d$ and $TPR_d$ by $\delta_1$ and $\delta_2$ for RF and NN can be seen in Fig. \ref{fignpvppvRF} and Fig. \ref{fignpvppvLM} respectively.

\noindent The lighter lines represent the $TPR_d$ and the darker lines $TNR_d$. The dotted lines are the lower limits for $TPR_d$ and $TNR_d$ respectively. We can see that the RF model achieves higher $TNR_d$, however, lower $TPR_d$. The step-form of the NN $TNR_d$ curve can be explained by very few predictions having probabilities over $0.5$, therefore, a small change in the thresholds does not have much impact.

\begin{figure}[h!]
\centerline{\includegraphics[width=90mm]{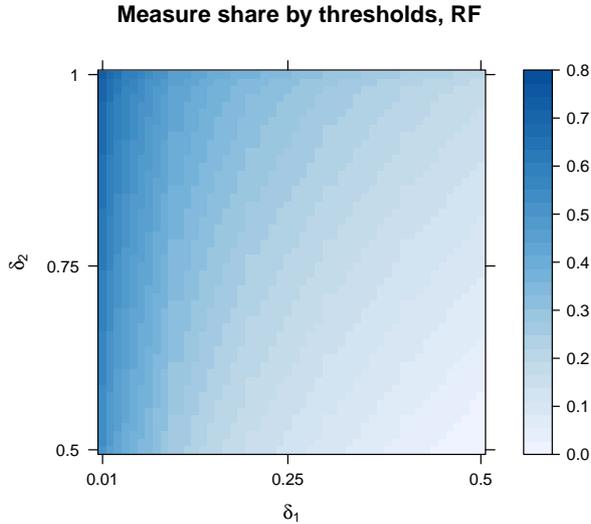}}
\caption{Measure share for Random Forest by $\delta_1$ and $\delta_2$.}
\label{figmeasureRF}
\end{figure}

\begin{figure}[h!]
\centerline{\includegraphics[width=90mm]{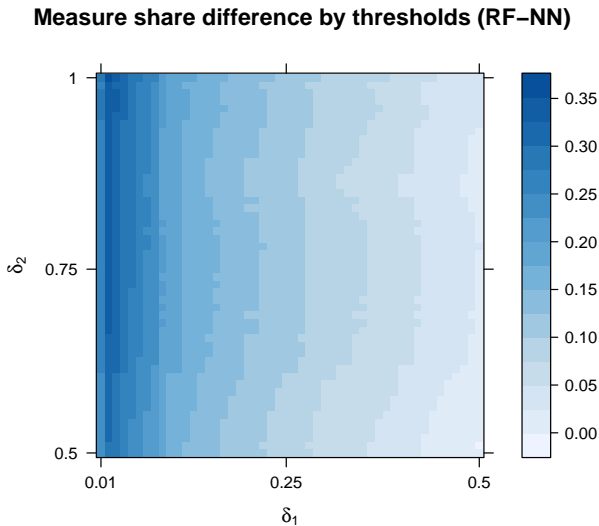}}
\caption{Measure share difference between Random Forest and Neural Networks by $\delta_1$ and $\delta_2$.}
\label{figmeasureLM}
\end{figure}

\indent The measure share by $\delta_1$ and $\delta_2$ can be seen in Fig. \ref{figmeasureRF} for RF. Fig. \ref{figmeasureLM} displays the difference in measure share between RF and NN.

\noindent The color scale to the right of the heatmaps shows the corresponding measure share. Each composition of $\delta_1$ and $\delta_2$ is represented by a coordinate in the map. We can see that the shade gradient in Fig. \ref{figmeasureRF} varies with the thresholds: for high values of $\delta_1$ and low values of $\delta_2$ the color shift is less prominent while the measure share shifts more rapidly for low values of $\delta_1$ and high values of $\delta_2$. In the difference plot, the darker the shade, the better the performance of NN in comparison to RF. The NN seem to outperform the RF for low values of $\delta_1$ in terms of measure share. Note that both thresholds have a prominent effect on the share of observations recommended for measuring. We define the best model as the model with the lowest measure share, yet meeting the lower limits of $TPR_d=0.90$ and $TNR_d=0.95$. The best results for each model are shown in Tab. \ref{tab2}.

\begin{table}[h!]
\caption{Best thresholds}
\begin{center}
\begin{tabular}{lccccc} \hline
\textbf{Model} & $\boldsymbol{\delta_1}$ & $\boldsymbol{\delta_2}$ & $\mathbf{TPR_d}$ & $\mathbf{TNR_d}$ & \textbf{M. share}\\\hline
\textbf{RF} & 0.50 & 0.65 & 0.919 & 0.952 & 0.0563\\
\textbf{NN} & 0.50 & 0.95 & 0.985 & 0.952 & 0.138\\\hline
\end{tabular}
\label{tab2}
\end{center}
\end{table}

\noindent We can see that both models fulfill the lower limits. The Neural Networks achieve higher $TPR_d$ than Random Forest, however, Random Forest allows a higher $\delta_1$ and a lower $\delta_2$, still maintaining a lower share of observations which has to be measured. Despite the indications in Fig. \ref{figmeasureLM} of NN achieving a lower measure share than RF, the NN requires a higher $\delta_2$ to reach the lower limit of $TNR_d$ which results in the best NN model having higher measure share than the best RF model. Therefore, the Random Forest is, according to the results, superior to the Neural Networks.

\section{Conclusions}
An easy yet costly approach to obtain inter-frequency information is to perform frequent measuring. Our results show that standard machine learning methods perform unsatisfactory, even using oversampling to counter imbalanced classes.
Our duo-threshold approach combines a highly confident prediction phase with a measurement phase. This allows for a high accuracy of $92\%$ and $95\%$ for two types of decisions while reducing the need for inter-frequency measuring by approximately $95\%$ in comparison to frequent measuring. The authentic data is combined with some limitations in information which could be of importance, such as lack of geographical position and cell load. The models could also be improved if measurements from several alternative frequencies for each observation were available.

\section*{Acknowledgments}
This work was partially supported by the Wallenberg AI, Autonomous Systems and Software Program (WASP) funded by the Knut and Alice Wallenberg Foundation. We are also grateful to Martin Isaksson, Rickard C\"oster, Samuel Axelsson and Vijaya Yajnanarayana for their detailed comments.



\begin{thebibliography}{00}
\bibitem{handover} Dimou, K., Wang, M., Yang, Y., Kazmi, M., Larmo, A., Pettersson, J., Muller, W. and Timner, Y., Handover within 3GPP LTE: Design Principles and Performance, IEEE 70th Vehicular Technology Conference Fall, 2009.
\bibitem{rsrp} Ryd\'en, H., Berglund, J., Isaksson, M., C\"oster, R. and Gunnarsson, F., Predicting Strongest Cell on Secondary Carrier
using Primary Carrier Data, IEEE Wireless Communications and Networking Conference Workshops (WCNCW): 7th International Workshop on Self-Organizing
Networks (IWSON), pp. 137--142, 2018.
\bibitem{rsrq} Evolved Universal Terrestrial Radio Access (E-UTRA),''Physical layer; Measurements'' in Technical Specification (TS) 36.214, 3rd Generation Partnership Project (3GPP), 2018. URL: \href{http://www.3gpp.org/ftp//Specs/archive/36_series/36.214/}{http://www.3gpp.org/ftp//Specs/archive/ \\ 36\_series/36.214/}.
\bibitem{forest} Breiman, L., Random Forests, Machine Learning, vol. 45(1), 2001, pp. 5--32.
\bibitem{boot} Efron, B., Bootstrap methods: another look at the jackknife, Annals of Statistics, 7, pp. 1--26, 1979.

\bibitem{bishop} C.M. Bishop, ``Pattern recognition and machine learning'', Springer, 12, pp. 138--139, 205--206, 227-228, 304--305, 666, 2006.

\bibitem{R} R Core Team, ''R: A language and environment for statistical computing''. R Foundation for Statistical Computing, Vienna, Austria, 2017. URL: \href{https://www.R-project.org/}{https:\/\/www.R-project.org\/}.
\bibitem{rForest} Liaw, A. and Wiener, M., ''Classification and Regression by randomForest'', R News 2(3), 2002, pp. 18--22. 
\bibitem{rGP} Karatzoglou, A.,  Smola, A., Hornik, K., and Zeileis, A. ''kernlab - An S4 Package for Kernel Methods in R'', Journal of Statistical Software 11(9), 2004, pp. 1--20. 
\bibitem{nn} Fritsch, S. and Guenther, F., ''neuralnet: Training of Neural Networks'', 2016. URL: \href{https://CRAN.R-project.org/package=neuralnet}{https://CRAN.R-project.org/package=neuralnet}

\end{thebibliography}
\end{document}